\documentclass[pra,a4paper,showpacs,superscriptaddress]{revtex4-2}
\usepackage{amsmath}
\usepackage{amsfonts}
\usepackage{graphicx}
\usepackage{longtable}
\newcommand{\be}{\begin{equation}}
\newcommand{\ee}{\end{equation}}
\newcommand{\bea}{\begin{eqnarray}}
\newcommand{\eea}{\end{eqnarray}}

\newcommand{\ba}[1]{\left(\begin{array}{#1}}
\newcommand{\ea}{\end{array}\right)}
\begin{document}

\title{Canonical steering ellipsoids of pure symmetric multiqubit states with 
two distinct spinors and volume monogamy of steering} 
\author{B. G. Divyamani} 
\affiliation{Tunga Mahavidyalaya, Thirthahalli-577432, Karnataka, India}
\author{I. Reena} 
\affiliation{Department of Physics, Bangalore University, Bangalore-560 056, 
India} 
\author{Prasanta K. Panigrahi}
\affiliation{Department of Physical Sciences, Indian Institute of Science Education
and Research Kolkata, Mohanpur-741246, West Bengal, India} 
\author{A. R. Usha Devi}
\affiliation{Department of Physics, Bangalore University, 
Bangalore-560 056, India} 
\affiliation{Inspire Institute Inc., Alexandria, Virginia, 22303, USA.}
\author{Sudha} 
\email{tthdrs@gmail.com} 
\affiliation{Department of Physics, Kuvempu University, 
Shankaraghatta-577 451, Karnataka, India}
\affiliation{Inspire Institute Inc., Alexandria, Virginia, 22303, USA.}
\date{\today}

\begin{abstract} 
Quantum steering ellipsoid formalism provides a faithful
representation of all two-qubit states and is useful in obtaining their correlation properties. The steering ellipsoids of two-qubit states that have undergone local operations on both the qubits so as to bring the state to its canonical form are the so-called {\emph{canonical steering ellipsoids}}. The steering ellipsoids corresponding to the two-qubit subsystems of permutation symmetric $N$-qubit states are considered here. 
We construct and analyze the geometric features of the canonical steering ellipsoids corresponding to pure permutation symmetric $N$-qubit states with two distinct spinors. Depending on the degeneracy of the two spinors in the pure symmetric $N$-qubit state, there arise several families which cannot be converted into one another through Stochastic Local Operations and Classical Communications (SLOCC).  The canonical steering ellipsoids of the two-qubit states drawn from the pure symmetric $N$-qubit states with two distinct spinors allow for a geometric visualization of the SLOCC equivalent class of states. We show that the states belonging to the W-class correspond to oblate spheroids centered at $(0,0,1/(N-1))$ with fixed semiaxes lengths 
$1/\sqrt{N-1}$ and $1/(N-1)$. The states belonging to all other SLOCC inequivalent families 
correspond to ellipsoids centered at the origin of the Bloch sphere. We also explore volume monogamy relations of states belonging to these families, mainly the W-class of states. 
\end{abstract}
\pacs{03.65.Ud, 03.67.Bg} 
\maketitle

\section{Introduction} 
The Bloch sphere representation of a single qubit contains  valuable geometric information needed for quantum information processing tasks.   
A natural generalization and an analogous picture for a two-qubit system is provided by the {\emph{quantum steering ellipsoid}}~\cite{jevtic2014,MilneNJP2014,MilnePRA2016} and is helpful in understanding correlation properties such as quantum discord~\cite{shi2011,shi}, volume monogamy of steering~\cite{MilneNJP2014,MilnePRA2016} etc.,  Quantum steering ellipsoid is the set of all Bloch vectors to which one party's qubit could be `steered' when 
all possible measurements are carried out on the qubit belonging to other party. The volume of the steering ellipsoids~\cite{jevtic2014} corresponding to the two-qubit subsystems of an $N$-qubit state, $N>3$, capture monogamy properties of the state  effectively~\cite{MilneNJP2014,MilnePRA2016} and provides insightful information about two-qubit entanglement.   

While the quantum steering ellipsoid~\cite{jevtic2014,MilneNJP2014,MilnePRA2016} is the set of all Bloch vectors of first qubit steered by local operations on second qubit, the so-called {\emph{canonical steering ellipsoid}}~\cite{verstraete2001,fvthesis,supra} is the steering ellipsoid of a two-qubit state that has attained a canonical form under suitable SLOCC operations on {\emph {both the qubits}}. It has been shown that the SLOCC canonical forms of a two-qubit state can either be a Bell diagonal form or a nondiagonal one (when the two-qubit state is rank-deficient)~\cite{verstraete2001,supra}. The canonical steering ellipsoids corresponding to the two-qubit states can thus have only two distinct forms~\cite{verstraete2001,supra} and  provide a much simpler geometric picture representing the set of all SLOCC equivalent two-qubit states. 

The canonical steering ellipsoids corresponding to the two-qubit subsystems of pure three-qubit permutation symmetric states are analyzed in Ref.~\cite{can}. It has been shown that~\cite{can} the two SLOCC inequivalent families of  pure three-qubit permutation symmetric states, the W-class of states (with two distinct spinors) and the GHZ class of states (with three distinct spinors)  correspond to distinct canonical steering ellipsoids. While an ellipsoid centered at the origin of the Bloch sphere is the canonical steering ellipsoid for the GHZ class of states, an oblate spheroid with its center shifted along the polar axis is the one for W-class of states. Using these, the volume monogamy relations are established and the obesity of the steering ellipsoids is made use of to obtain expressions for concurrence of states belonging to these two SLOCC inequivalent families in Ref.~\cite{can}. 

In this paper, we extend the analysis to a class of 
$N$-qubit pure states which are symmetric under exchange of qubits.  Through the SLOCC canonical forms of the two-qubit reduced state, extracted from pure symmetric {\emph{multiqubit} states with two distinct spinors and the Lorentz canonical forms of their real representative, we examine the features of {\emph {canonical steering ellipsoids}} associated with them. We identify the special features of the canonical steering ellipsoid representing $N$-qubit states of the W-class and these features distinguish this class from all other SLOCC inequivalent families of pure symmetric $N$-qubit states.  We discuss the volume monogamy of steering for pure permutation symmetric $N$-qubit states and obtain the volume monogamy relation satisfied by W-class of states. An expression for obesity of the steering ellipsoid and thereby an expression for concurrence of two-qubit subsystems of $N$-qubit states belonging to the W-class is obtained.  

Contents of this paper are organized as follows: In Sec.II, we give a brief review on SLOCC classification of pure permutation symmetric multiqubit states based on Majorana  representation~\cite{majorana,bastin,solano,arus} and obtain the two-qubit subsystems of the states belonging to SLOCC inequivalent families of pure symmetric multiqubit states with two distinct spinors. Sec.~III provides an outline of the real matrix representation of a two-qubit density matrix and their Lorentz canonical forms under SLOCC transformation of the two-qubit density matrix. We also obtain the Lorentz canonical forms of two-qubit subsystems corresponding to SLOCC inequivalent families, in Sec.~III. In Sec.IV, we analyse the nature of steering ellipsoids associated with the distinct Lorentz canonical forms obtained in Sec.~III. The volume monogamy of steering for pure symmetric multiqubit states with two distinct spinors is discussed along with illustration for W-class of states, in Sec.~V. Summary of our results is presented in Sec.~VI.   
    
\section{Majorana geometric representation of pure symmetric $N$-qubit states with two distinct spinors} 

Ettore Majorana, in his novel 1932 paper~\cite{majorana} proposed that a pure spin $j=\frac{N}{2}$ quantum state can  be represented as a {\em symmetrized} combination of $N$ constituent spinors as follows:
\begin{equation}
\label{Maj}
\vert \Psi_{\rm sym}\rangle={\mathcal N}\, \sum_{P}\, \hat{P}\, \{\vert \epsilon_1, \epsilon_2, 
\ldots  \epsilon_N \rangle\}, 
\end{equation}  
where 
\begin{equation}
\label{spinor}
\vert\epsilon_l\rangle= \left(
\cos(\alpha_l/2)\, \vert 0\rangle +
\sin(\alpha_l/2) \, \vert 1\rangle\right) e^{i\beta_l/2},\ \ l=1,\,2,\ldots,\,N.
\end{equation}
The symbol $\hat{P}$ corresponds to the set of all $N!$ permutations of the spinors (qubits) and ${\mathcal N}$ corresponds to an overall normalization factor. 
The name Majorana {\emph {geometric}} representation is owing to the fact that it leads to an intrinsic geometric picture of the  state in terms of $N$ points on the unit sphere. In fact, the spinors $\vert \epsilon_l\rangle$, $l=1,\,2,\ldots,\,N$ of (\ref{spinor}) correspond geometrically to $N$ points on the unit sphere $S^2$, with the pair of angles $(\alpha_l,\beta_l)$ determining the orientation of each point on the sphere. 

The pure symmetric $N$-qubit states characterized by {\emph{two}} distinct qubits  
are given by~\cite{bastin,solano,arus}, 
\begin{eqnarray}
\label{dnk}
\vert D_{N-k, k}\rangle &=& {\mathcal N}\, \sum_{P}\, \hat{P}\,\{ \vert \underbrace{\epsilon_1, \epsilon_1,
\ldots , \epsilon_1}_{N-k};\ \underbrace{\epsilon_2, \epsilon_2,\ldots, \epsilon_2}_{k}\rangle\}.
\end{eqnarray} 
Here,  one of the spinors say $\vert \epsilon_1 \rangle$ occurs $N-k$ times whereas the  other spinor $\vert \epsilon_2 \rangle$ occurs $k$ times in each term of the symmetrized combination. 
Under identical local unitary transformations, the pure symmetric $N$-qubit states with two distinct spinors 
can be brought  to the canonical  form~\cite{arus}, 
\begin{eqnarray}
\label{nono}
\vert D_{N-k, k}\rangle &\equiv & \sum_{r=0}^k\, \beta^{(k)}_{r}\,\,  \left\vert\frac{N}{2},\frac{N}{2}-r \right\rangle,  \ \ \ \ k=1,\,2,\,3,\ldots \left[\frac{N}{2}\right]\\ 
\label{nono1}
\beta^{(k)}_{r}&=&{\mathcal N}\,\, 
\sqrt{\frac{N!(N-r)!}{r!}}\,\frac{a^{k-r}\, b^r}{(N-k)! (k-r)!},\ \ \ \ 0\leq a<1, \ \ b=\sqrt{1-a^2}. 
\end{eqnarray} 
Notice that $\left\vert\frac{N}{2},\frac{N}{2}-r \right\rangle$, $r=0,\,1,\,2\ldots,$ are the Dicke states, which are common eigenstates of collective angular momentum operators $J^2$ and $J_z$. They are basis states of the $N+1$ dimensional symmetric subspace of collective angular momentum space of $N$ qubits. 
The states $\vert D_{N-k, k}\rangle$ (see (\ref{nono}), (\ref{nono1})) are characterized by only one real parameter `$a$' and thus form one parameter family of states 
$\{{\mathcal{D}}_{N-k,k}\}$~\cite{arus,saruakr}.  
When $a=0$, the states $\vert D_{N-k, k}\rangle$ reduce to the Dicke states $\left\vert N/2,\,N/2-k \right\rangle$~\cite{arus,saruakr} in which $\vert\epsilon_1\rangle=\vert 0\rangle$ and $\vert \epsilon_2\rangle=\vert 1\rangle$ (see (\ref{dnk})). When $a\longrightarrow 1$, $\vert D_{N-k, k}\rangle$ becomes a separable state 
consisting of only one spinor $\vert \epsilon_1\rangle$ or $\vert \epsilon_2\rangle$.   

It is important to notice that in the family $\{{\mathcal{D}}_{N-k,k}\}$, different values of $k$, ($k=1,\,2,\,3,\ldots \left[\frac{N}{2}\right]$), correspond to different  SLOCC inequivalent classes~\cite{arus}. That is, a state $\vert D_{N-k, k}\rangle$ cannot be converted into 
$\vert D_{N-k',k'}\rangle$, $k\neq k'$ through any choice of local unitary (identical) transformations. In fact, different values of $k$ lead to different {\emph {degeneracy configurations}}~\cite{arus} of the two spinors $\vert \epsilon_1 \rangle$, $\vert \epsilon_2 \rangle$ in the state $\vert D_{N-k, k}\rangle$. When $k=1$, one gets  the W-class of states $\{{\mathcal{D}}_{N-1,1}\}$  where one of the qubits say $\vert \epsilon_1 \rangle$ repeats only once in each term of the symmetrized combination (see (\ref{dnk})) and the other qubit $\vert \epsilon_2 \rangle$ repeats $N-1$ times. The N-qubit W-state 
\begin{equation*} 
\vert {{W}}_N\rangle=\frac{1}{\sqrt{N}}\left[\vert 000\ldots 1\rangle+\vert 000\ldots 10\rangle+\cdots+\vert 100\ldots 00\rangle\right]\equiv \left\vert \frac{N}{2},\frac{N}{2}-1\right\rangle
\end{equation*}
belongs to the family $\{{\mathcal{D}}_{N-1,1}\}$ and hence the name {\emph{W-class of states}}. The Dicke state 
\begin{equation*} 
\left\vert \frac{N}{2},\frac{N}{2}-2\right\rangle=\sqrt{\frac{2}{N(N-1)}}\left[\vert 000\ldots011\rangle+\vert 000\ldots 0110\rangle+\cdots+\vert 110\ldots 00\rangle\right].
\end{equation*}
is a typical state of the family $\{{\mathcal{D}}_{N-2,2}\}$. 
In all, there are $\left[\frac{N}{2}\right]$ SLOCC inequivalent families in the set of all pure permutation symmetric $N$-qubit states with two-distinct spinors~\cite{fn}. 

\subsection{ Two-qubit reduced density matrices of the states $\vert D_{N-k,\,k}\rangle$} 
 The two-qubit marginal $\rho^{(k)}$ corresponding to any random pair of qubits in the pure symmetric  $N$-qubit state $\vert D_{N-k,\,k}\rangle\in 
\{{\mathcal{D}}_{N-k,k}\}$ is obtained by tracing over the remaining $N-2$ qubits in it. In Ref. \cite{akhiss}, it has been shown, using the algebra of addition of angular momenta, $j_1=1$ (corresponding to two-qubit marginal) and $j_2=(N-2)/2$ (corresponding to the remaining $N-2$ qubits), that the two-qubit reduced density matrix $\rho^{(k)}$ 
has the form 
\begin{eqnarray}
\label{rhok_matrix}
\rho^{(k)}&=&\ba{cccc} A^{(k)} \ \ & B^{(k)} \ \ & B^{(k)}\ \  & C^{(k)} \ \  \\ B^{(k)} \ \  & D^{(k)}\ \  & D^{(k)}\ \  & E^{(k)} \ \   \\ B^{(k)} \ \  & D^{(k)} \ \  & D^{(k)} \ \ & E^{(k)} \ \  \\ C^{(k)} \ \  & E^{(k)} \ \  & E^{(k)} \ \  & F^{(k)} \ \  \ea. 
\end{eqnarray}
The elements $A^{(k)},\, B^{(k)},\, C^{(k)},\, D^{(k)},\, E^{(k)}$ and $F^{(k)}$ are real and are explicitly given by~\cite{akhiss}
\begin{eqnarray}
\label{elements}
A^{(k)}=\sum_{r=0}^k\, \left({\beta_r^{k}}\right)^2 \left({c^{(r)}_{1}}\right)^2, \  & & \  B^{(k)}=\frac{1}{\sqrt{2}}\sum_{r=0}^{k-1}\, 
{\beta^{(k)}_r} \beta^{(k)}_{r+1}\, c^{(r)}_{1} 
c^{(r+1)}_{0} \nonumber \\ 
& & \nonumber \\ 
C^{(k)}=\sum_{r=0}^{k-2}\, \beta^{(k)}_r \beta^{(k)}_{r+2}\,\, c^{(r)}_{1} c^{(r+2)}_{-1}, \  & & \  
D^{(k)}=\frac{1}{2}\sum_{r=1}^{k}\, \left({\beta_r^{(k)}}\right)^2  \left({c^{(r)}_{0}}\right)^2 \\
& & \nonumber \\ 
E^{(k)}=\frac{1}{\sqrt{2}}\sum_{r=0}^{k-1}\, \beta^{(k)}_r \beta^{(k)}_{r+1}\,\, c^{(r)}_{0}c^{(r+1)}_{-1}, \  & &  \  
 \ \ \ F^{(k)}=\sum_{r=0}^k\, \left({\beta_r^{(k)}}\right)^2  \left({c^{(r)}_{-1}}\right)^2.  \nonumber
\end{eqnarray} 
where, $\beta_r^{(k)}$ are given as functions of the parameter `$a$' in (\ref{nono1}) and 
\begin{eqnarray}
\label{cg_explicit}
c^{(r)}_{1}&=&\sqrt{\frac{(N-r)(N-r-1)}{N(N-1)}},\ \ \ c^{(r)}_{-1}=\sqrt{\frac{r\, (r-1)}{N(N-1)}},\nonumber \\
c^{(r)}_{0}&=&\sqrt{\frac{2r\, (N-r)}{N(N-1)}}	  
\end{eqnarray} 
are the Clebsch-Gordan coefficients $c^{(r)}_{m_2}~=~C\left(\frac{N}{2}-1,\, 1,\, \frac{N}{2};m-m_2,\, m_2, m \right)$, $m~=~\frac{N}{2}-r$, $m_2=1,\,0,\,-1$ ~\cite{Var}. In particular, for W-class of states i.e., when $k=1$, we have 
\begin{eqnarray}
\label{rho1}
&&\rho^{(1)}=\mbox{Tr}_{N-2}\left(\vert D_{N-1,\,1}\rangle\langle D_{N-1,\,1}\vert\right)\nonumber \\
&&\ = \left( \left(\beta^{(1)}_0\right)^2+ \left(\beta^{(1)}_1\,  c^{(1)}_1\right)^2   \right)\vert 1,\,1\rangle\langle 1,\,1 \vert \nonumber \\ 
&& \ \ \  + \left(\beta^{(1)}_1\, c^{(1)}_0\right)^2  
\vert 1,\,0\rangle\langle 1,\,0 \vert  +\beta^{(1)}_0 \beta^{(1)}_1\, c^{(1)}_0 \vert 1,\,1\rangle\langle 1,\,0 \vert \nonumber \\ 
&&\ \ \  +\beta^{(1)}_0 \beta^{(1)}_1\, c^{(1)}_0 \vert 1,\,0\rangle\langle 1,\,1 \vert 
\end{eqnarray}
Here  (see (\ref{nono1})) we have $\beta^{(1)}_0={\mathcal N}N\, a $, $\beta^{(1)}_1={\mathcal N}\, \sqrt{N(1- a^2)}$ with ${\mathcal N}=\frac{1}{\sqrt{N^2\,a^2+N(1-a^2)}}$ and  the associated non-zero Clebsch-Gordan coefficients (see (\ref{cg_explicit})) are given by 
 \begin{equation}
 \label{cg_explicit1}
 c^{(1)}_1=\sqrt{\frac{N-2}{N}},\ \ \  c^{(1)}_0=\sqrt{\frac{2}{N}}.
\end{equation}
In the standard two-qubit basis $\{\vert 0_A,0_B \rangle, \vert 0_A,1_B \rangle, \vert 1_A,0_B \rangle, \vert 1_A,1_B \rangle\}$, the two-qubit density matrix  
$\rho^{(1)}$ drawn from the states $\vert D_{N-1,1}\rangle$ takes the form 
\begin{eqnarray}
\label{rho1_matrix}
\rho^{(1)}&=&\ba{cccc} A^{(1)} \ \ & B^{(1)} \ \ & B^{(1)}\ \  & 0 \ \  \\ B^{(1)} \ \  & D^{(1)}\ \  & D^{(1)}\ \  & 0 \ \   \\ B^{(1)} \ \  
& D^{(1)} \ \  & D^{(1)} \ \ & 0 \ \  \\ 0 \ \  & 0 \ \  & 0 \ \  & 0 \ \  \ea 
\end{eqnarray}
where 
\begin{eqnarray}
\label{rho1ele}
A^{(1)}&=&\frac{N^2a^2+(N-2)(1-a^2)}{N^2\,a^2+N(1-a^2)},\ \ B^{(1)}=\frac{a\sqrt{1-a^2}}{1+a^2(N-1)},\   \nonumber \\
 D^{(1)}&=& \frac{1-a^2}{N^2\,a^2+N(1-a^2)},\ 
 \end{eqnarray} 
In a similar manner, the two-qubit subsystems of pure symmetric $N$-qubit states $\vert D_{N-k,k}\rangle$ belonging to each SLOCC inequivalent family 
$\{{\mathcal D}_{N-k,\, k}\}$, $k=2,\,3,\ldots,\,\left[\frac{N}{2}\right]$ can be obtained as a function of $N$ and `$a$' using Eqs. (\ref{rhok_matrix}), (\ref{elements}), (\ref{cg_explicit}). As is shown in  Refs.~\cite{supra,can}, the real representative $\Lambda^{(k)}$ of the two-qubit subsystem $\rho^{(k)}$ and its Lorentz canonical form 
$\widetilde{\Lambda}^{(k)}$ are   essential in obtaining the geometric representation of the states $\vert D_{N-k,k}\rangle$, for all $k$. 
We thus proceed to obtain $\Lambda^{(k)}$ and its Lorentz canonical form $\widetilde{\Lambda}^{(k)}$ in the following. 

\section{The real representation of $\rho^{(k)}$ and its Lorentz canonical forms} 

The real representative $\Lambda^{(k)}$ of the two-qubit state $\rho^{(k)}$ is a $4\times 4$ real matrix with its elements given 
by 
\begin{eqnarray}
	\label{lambda}
	\Lambda^{(k)}_{\mu \, \nu}&=& {\rm Tr}\,\left[\rho^{(k)}\,
	(\sigma_\mu\otimes\sigma_\nu)\,\right]  
\end{eqnarray} 
That is, $\Lambda^{(k)}_{\mu \, \nu}$, $\mu, \nu=0,\,1,\,2,\,3$ are the coefficients of expansion of $\rho^{(k)}$, expanded in the Hilbert-Schmidt basis $\{\sigma_\mu\otimes \sigma_\nu\}$: 
\begin{eqnarray}
	\label{rho2q}
	\rho^{(k)}&=&\frac{1}{4}\, \sum_{\mu,\,\nu=0}^{3}\,   
	\Lambda^{(k)}_{\mu \, \nu}\, \left( \sigma_\mu\otimes\sigma_\nu \right), 
\end{eqnarray}
Here, $\sigma_i$, $i=1,\,2,\,3$ are the Pauli spin matrices and $\sigma_0$ is the $2\times 2$ identity matrix;
\begin{eqnarray}
	\label{sigmamu}
	\sigma_0=\left(\begin{array}{cc} 1 & 0 \\ 0 & 1       \end{array}\right),\ \  \sigma_1=\left(\begin{array}{cc} 0 & 1 \\ 1 & 0       \end{array}\right),\ \   \sigma_2=\left(\begin{array}{cc} 0 & -i \\ i & 0       \end{array}\right),\ \   \sigma_3=\left(\begin{array}{cc} 1 & 0 \\ 0 & -1  \end{array}\right).   
\end{eqnarray} 
It can be readily seen that (see (\ref{lambda}), (\ref{rho2q})) the real $4\times 4$ matrix $\Lambda^{(k)}$ has the form  
\begin{eqnarray}
\label{Lg}
	\Lambda^{(k)}&=&\left(\begin{array}{llll} 1 & r_1 & r_2 & r_3 \\ 
		s_1 & t_{11}  & t_{12} & t_{13} \\ 
		s_2 & t_{21}  & t_{22} & t_{23} \\
		s_3 & t_{31}  & t_{32} & t_{33} \\
	\end{array}\right),  
\end{eqnarray} 
where ${\mathbf r}=(r_1,\,r_2,\,r_3)^T$, ${\mathbf s}=(s_1,\,s_2,\,s_3)^T$ are Bloch vectors of the individual qubits and $T=(t_{ij})$ is the correlation matrix;
\begin{eqnarray} 
	\label{ri}
	r_i&=&\Lambda^{(k)}_{i \, 0}= {\rm Tr}\,\left[\rho^{(k)}\, (\sigma_i\otimes\sigma_0)\,\right] \ \ \\
	\label{sj} 
	s_j&=& \Lambda^{(k)}_{0 \, j}={\rm Tr}\,\left[\rho^{(k)}\, (\sigma_0\otimes\sigma_j)\,\right] \\ 
	\label{tij}
	t_{ij}&=& \Lambda^{(k)}_{i \, j}= {\rm Tr}\,\left[\rho^{(k)}\, (\sigma_i\otimes\sigma_j)\,\right],\  \ \ \ i,\,j=1,\,2,\,3. 
\end{eqnarray} 
For a symmetric two-qubit density matrix, the Bloch vectors $\mathbf{r}$ and $\mathbf{s}$ are identical and hence $r_i=s_i$, $i=1,\,2,\,3$;  From the  structure of $\rho^{(k)}$ in (\ref{rhok_matrix}) and using (\ref{ri}), (\ref{sj}), (\ref{tij}) we obtain the general form of the real matrix $\Lambda^{(k)}$ as
\begin{eqnarray}
\label{Lk}
	\Lambda^{(k)}&=&\left(\begin{array}{cccc} 1& \frac{2(B^{(k)}+E^{(k)})}{A^{(k)}+2D^{(k)}+F^{(k)}} & 0 & \frac{A^{(k)}-F^{(k)}}{A^{(k)}+2D^{(k)}+F^{(k)}} \\ 
		\frac{2(B^{(k)}+E^{(k)})}{A^{(k)}+2D^{(k)}+F^{(k)}} & \frac{2(C^{(k)}+D^{(k)})}{A^{(k)}+2D^{(k)}+F^{(k)}}  & 0 & \frac{2(B^{(k)}-E^{(k)})}{A^{(k)}+2D^{(k)}+F^{(k)}} \\ 
		0 & 0  & \frac{2(D^{(k)}-C^{(k)})}{A^{(k)}+2D^{(k)}+F^{(k)}} & 0 \\
		\frac{A^{(k)}-F^{(k)}}{A^{(k)}+2D^{(k)}+F^{(k)}}  & \frac{2(B^{(k)}-E^{(k)})}{A^{(k)}+2D^{(k)}+F^{(k)}}  & 0 & 1-\frac{4D^{(k)}}{A^{(k)}+2D^{(k)}+F^{(k)}} \\
	\end{array}\right).  
\end{eqnarray} 
The elements of $\Lambda^{(k)}$, for different $k$, can be evaluated using (\ref{elements}), (\ref{cg_explicit})):

\subsection{Lorentz canonical forms of $\Lambda^{(k)}$}

Under SLOCC transformation, the two-qubit density matrix $\rho^{(k)}$ transforms to $\widetilde{\rho}^{(k)}$ as  
	\begin{eqnarray}
	\label{rhokab}
		\rho^{(k)}\longrightarrow\widetilde{\rho}^{(k)}&=&\frac{(A\otimes B)\, \rho^{(k)}\, (A^\dag\otimes B^\dag)}
		{{\rm Tr}\left[\rho^{(k)}\, (A^\dag\, A\otimes B^\dag\, B)\right]}.
	\end{eqnarray} 
	Here, $A, B\in {\rm SL(2,C)}$ denote $2\times 2$ complex matrices with unit determinant. A suitable choice of $A$ and $B$ takes the two-qubit density matrix $\rho^{(k)}$ to its canonical form $\widetilde{\rho}^{(k)}$.
	
	The transformation of $\rho^{(k)}$  in (\ref{rhokab}) leads to the transfomation~\cite{supra,can} 
	\begin{eqnarray}
		\label{sl2c}
		\Lambda^{(k)}\longrightarrow \widetilde{\Lambda}^{(k)}&=&\frac{L_A\,\Lambda^{(k)}\, L^T_B}{\left(L_A\,\Lambda^{(k)}\, L^T_B\right)_{00}}.
	\end{eqnarray} 
	of   its real representative 
	$\Lambda^{(k)}$.
	In (\ref{sl2c}), $L_A,\, L_B\in SO(3,1)$ are $4\times 4$  proper orthochronous Lorentz transformation matrices~\cite{KNS} corresponding respectively to  $A$, 
	$B\in SL(2,C)$  and the superscript `$T$' denotes transpose operation.  
	The Lorentz canonical form $\widetilde{\Lambda}^{(k)}$ of $\Lambda^{(k)}$ and thereby the SLOCC canonical form of the two-qubit density matrix $\rho^{(k)}$ (see (\ref{rhokab})) can be obtained by constructing the $4\times 4$ real symmetric matrix $\Omega^{(k)}=\Lambda^{(k)}\, G\, \left(\Lambda^{(k)}\right)^T$, where $G={\rm diag}\,(1,-1,-1,-1)$ denotes the Lorentz metric. Using the defining property~\cite{KNS}   $L^T\,G\,L=G$  of Lorentz transformation $L$,  it can be seen that $\Omega^{(k)}$
	undergoes a {\em Lorentz congruent transformation} under SLOCC (up to an overall  factor)~\cite{supra} as   
	\begin{eqnarray}
		\label{oa} 
		\Omega^{(k)}\rightarrow \widetilde{\Omega}^{(k)}_A&=& \widetilde{\Lambda}^{(k)}\, G\, \left(\widetilde{\Lambda}^{(k)}\right)^T \nonumber \\
		&=& L_{A}\, \Lambda^{(k)}\, L_{B}^T\, G \, L_{B}\, {\Lambda^{(k)}}^T L_{A}^T \nonumber \\ 
		&=& L_{A}\, \Omega^{(k)}\, L_{A}^T. 
	\end{eqnarray}
It has been shown in Ref.~\cite{supra} that $\widetilde{\Lambda}^{(k)}$ can either be a real $4\times 4$ diagonal matrix or a non-diagonal matrix with only one off-diagonal element, depending on the eigenvalues, eigenvectors of  $G\,\Omega^{(k)}=G\left(\Lambda^{(k)}\, G\, \left(\Lambda^{(k)}\right)^T\right)$.  
\begin{itemize}
	\item[(i)] The diagonal canonical form 
		$\widetilde{\Lambda}^{(k)}_{I_c}$ results when the eigenvector $X_0$ associated with the highest eigenvalue $\lambda_0$ of $G\,\Omega^{(k)}$ obeys the   Lorentz invariant condition
		$X_0^T\, G\, X_0>0$. The diagonal canonical form $\widetilde{\Lambda}^{(k)}_{I_c}$ is explicitly given by
			\begin{eqnarray} 
			\label{lambda1c}
			\Lambda^{(k)}\longrightarrow\widetilde{\Lambda}^{(k)}_{I_c}&=&\frac{L_{A_1}\,\Lambda^{(k)}\, L^T_{B_1}}{\left(L_{A_1}\,\Lambda^{(k)}\, L^T_{B_1}\right)_{00}}\nonumber \\
			&=&{\rm diag}\,  \left(1,\,\sqrt{\frac{\lambda_1}{\lambda_0}},\sqrt{\frac{\lambda_2}{\lambda_0}},\, \pm\, \sqrt{\frac{\lambda_3}{\lambda_0}}\right), 
			\end{eqnarray} 
		where $\lambda_0\geq\lambda_1\geq\lambda_2\geq \lambda_3> 0$ are the  {\em non-negative} eigenvalues of $G\,\Omega^{(k)}$. 
		 The Lorentz transformations $L_{A_1},\, L_{B_1}\in SO(3,1)$ in (\ref{lambda1c}) respectively correspond to $SL(2,C)$ transformation matrices $A_1,\, B_1$  which take the two-qubit density matrix $\rho^{(k)}$ to its SLOCC canonical form $\widetilde{\rho}^{(k)}_{I_c}$ through the transformation (\ref{rhokab}). The diagonal form of $\widetilde{\Lambda}^{(k)}_{I_c}$ readily leads, on using (\ref{rho2q}), to Bell-diagonal form  
		\begin{eqnarray}	 
			\label{rhobd} 
			\widetilde{\rho}^{(k)}_{\,I_c}
			&=&  \frac{1}{4}\, \left( \sigma_0\otimes \sigma_0 + \sum_{i=1,2}\, \sqrt{\frac{\lambda_i}{\lambda_0}}\, \left(\sigma_i\otimes\sigma_i\right) \pm \sqrt{\frac{\lambda_3}{\lambda_0}}\, \left(\sigma_3\otimes\sigma_3\right) \right)
		\end{eqnarray}
		as the canonical form of the two-qubit state $\rho^{(k)}$.
		\item[(ii)]  The Lorentz canonical form of $\Lambda^{(k)}$ turns out to be a non-diagonal matrix (with only one non-diagonal element) given by  
		\begin{eqnarray} 
			\label{lambda2c}
			\Lambda^{(k)}\longrightarrow\widetilde{\Lambda}^{(k)}_{II_c}&=&\frac{L_{A_2}\,\Lambda^{(k)}\, L^T_{B_2}}{\left(L_{A_{2}}\, \Lambda^{(k)}\, L^T_{B_{2}}\right)_{00}} = \left(\begin{array}{cccc}
				1 & 0  & 0 & 0 \\ 
				0 & a_1 & 0 & 0 \\ 
				0 & 0 &   -a_1 & 0 \\
				1-a_0 & 0 & 0 &  a_0 \ \ 
			\end{array}\right) \ \ 
			\end{eqnarray}
	when the non-negative eigenvalues of $G\Omega^{(k)}$ are doubly degenerate with $\lambda_0\geq \lambda_1$ and the eigenvector $X_0$ belonging to the highest eigenvalue $\lambda_0$ satisfies the Lorentz invariant condition 	$X_0^T\, G\, X_0 =0$. In Ref.~\cite{supra}, it has been shown that when the maximum amongst the doubly degenerate eigenvalues of $G\Omega^{(k)}$ possesses an eigenvector $X_0$ satisfying the condition $X_0^T\, G\, X_0 =0$, the real symmetric matrix 
	$\Omega^{(k)}=\Lambda^{(k)}G\left(\Lambda^{(k)}\right)^T$ attains the non-diagonal Lorentz canonical form given by  
	\begin{eqnarray}
	\label{yyy}
	\Omega^{(k)}_{II_c}&=&\widetilde{\Lambda}^{(k)}_{II_c}\,G\,\left(\widetilde{\Lambda}^{(k)}_{II_c}\right)^T=L_{A_2}\,\Omega^{(k)}\, L^T_{A_2} \nonumber \\ 
	&=& \,\left(\begin{array}{cccc}
				\phi_0 & 0  & 0 & \phi_0-\lambda_0 \\ 
				0 & -\lambda_1 & 0 & 0 \\ 
				0 & 0 &   -\lambda_1 & 0 \\
				\phi_0-\lambda_0 & 0 & 0 &  \phi_0-2\lambda_0 \ \ 
			\end{array}\right).
	\end{eqnarray} 
The parameters $a_0$, $a_1$ in (\ref{lambda2c}) are related to the eigenvalues $\lambda_0$, $\lambda_1$ of $G\Omega^{(k)}$ and the   
$00^{\rm th}$ element of $\widetilde{\Omega}^{(k)}_{II_c}$ (see (\ref{yyy})). It can be seen that~\cite{supra} 
\begin{eqnarray}
\label{phi0}
&& a_0=\frac{\lambda_0}{\phi_0},\ \ a_1=\sqrt{\frac{\lambda_1}{\phi_0}},\ \  \mbox{where} \ \ \phi_0=\left(\Omega^{(k)}_{II_c}\right)_{00}=\left[\left(L_{A_2}\,\Lambda^{(k)}\, L^T_{B_2}\right)_{00}\right]^2.
\end{eqnarray} 
The Lorentz matrices $L_{A_{2}},\, L_{B_{2}}\in SO(3,1)$ correspond to  the SL(2,C) transformations $A_{2}$, $B_{2}$ that transform $\rho^{(k)}$ to its SLOCC canonical form $\rho^{(k)}_{II_c}$ (see \ref{rhokab}). The non-diagonl canonical form $\widetilde{\Lambda}^{(k)}_{II_c}$ leads to the SLOCC canonical form $\widetilde{\rho}^{(k)}_{\,II_c}$ of the two-qubit density matrix $\rho^{(k)}$, on using (\ref{rho2q});   
\begin{equation}
\label{rho2}
\widetilde{\rho}^{(k)}_{\,II_c}=\frac{1}{2}\left(\begin{array}{cccc}
				1 & 0  & 0 & a_1 \\ 
				0 & 1-a_0 & 0 & 0 \\ 
				0 & 0 &  0 & 0 \\
				a_1 & 0 & 0 &  a_0 \ \  
			\end{array}\right); \ \ \ 0\leq a_1^2\leq a_0 \leq 1.
\end{equation}
\end{itemize} 

\subsection{Lorentz canonical form of $\Lambda^{(1)}$ corresponding to W-class of states $\{{\mathcal{D}}_{N-k,k}\}$:} 
Using the explicit structure of the two-qubit state $\rho^{(1)}$ given in (\ref{rho1_matrix}), (\ref{rho1ele}), 
its real representative $\Lambda^{(1)}$ is obtained as (see (\ref{lambda}))   
\begin{eqnarray}
\label{lambda32}
\Lambda^{(1)}&=&\left(\begin{array}{cccc}1 & \frac{2a\sqrt{1-a^2}}{1+a^2(N-1)} &0& 1+\frac{2a^2}{1+a^2(N-1)}-\frac{2}{N}  \\ 
\frac{2a\sqrt{1-a^2}}{1+a^2(N-1)} & \frac{2(1-a^2)}{N\left(1+a^2(N-1)\right)} & 0  & \frac{2a\sqrt{1-a^2}}{1+a^2(N-1)} \\ 
0 & 0 &  \frac{2(1-a^2)}{N\left(1+a^2(N-1)\right)}  & 0 \\ 
1+\frac{2a^2}{1+a^2(N-1)}-\frac{2}{N} &  \frac{2a\sqrt{1-a^2}}{1+a^2(N-1)} & 0 & 1+\frac{4a^2}{1+a^2(N-1)}-\frac{4}{N}   \end{array} \right)= \left(\Lambda^{(1)}\right)^T. 
\end{eqnarray} 
We now construct the $4\times 4$ symmetric matrix $\Omega^{(1)}$ and obtain
\begin{eqnarray}
\label{omega32}
\Omega^{(1)}&=&\Lambda^{(1)}\, G\, \left(\Lambda^{(1)}\right)^T= \Lambda^{(1)}\, G\, \Lambda^{(1)}\nonumber \\ 
&=& \chi\left(\begin{array}{cccc} 
	N-1 & 0 &0 & N-2 \\
	0& -1 & 0  &  0 \\ 
	0 & 0 &  -1 & 0 \\ 
    N-2  &  0 & 0 & N-3\end{array} \right),\ \ \ \chi =\left[\frac{2(1-a^2)}{N\left(1+a^2(N-1)\right)}\right]^2. 
\end{eqnarray} 
The eigenvalues of the matrix $G\,\Omega^{(1)},\ G={\rm diag}\, (1,\,-1,\,-1,\,-1)$ are readily seen to be four-fold degenerate and are given by 
\begin{eqnarray}
	\label{ev32}
	\lambda_0&=&\lambda_1=\lambda_2=\lambda_3=\chi=\left[\frac{2(1-a^2)}{N\left(1+a^2(N-1)\right)}\right]^2. 
	\end{eqnarray}
It can be seen that $X_0=(1,\, 0,\, 0,\, -1)$ is an eigenvector of $G\,\Omega^{(1)}$  belonging to the four-fold degenerate eigenvalue $\lambda_0$ and   obeys the Lorentz invariant condition $X_0^T\, G\, X_0=0$. 
	We notice here that $\Omega^{(1)}$ is already in the canonical form (\ref{yyy}). On comparing (\ref{omega32}) with (\ref{yyy}), we get
	\begin{equation}
	\label{phi0g} 
	\phi_0=(\Omega^{(1)})_{00}=(N-1)\chi.
	\end{equation}
	On substituting the parameters $a_0$, $a_1$ (see (\ref{phi0}), (\ref{ev32}), (\ref{phi0g}))  in (\ref{lambda2c}), we arrive at the Lorentz canonical form  of the real matrix  $\Lambda^{(1)}$ as
 \begin{eqnarray}
 	\label{l32c}
 	\widetilde{\Lambda}^{(1)}&=&\left(\begin{array}{cccc}1 & 0 &0& 0  \\ 
 		0& \frac{1}{\sqrt{N-1}} & 0  &  0\\ 
 		0 & 0 &  -\frac{1}{\sqrt{N-1}}  & 0 \\ 
 		\frac{N-2}{N-1}  &  0 & 0 &  \frac{1}{N-1} \end{array} \right). 
 \end{eqnarray} 
 It can be readily seen that $\widetilde{\Lambda}^{(1)}$, the Lorentz canonical form corresponding to the W-class of states, is independent of the parameter `$a$'.

\subsection{Lorentz canonical form of $\Lambda^{(k)}$, $k=2,\,3,\ldots,\left[\frac{N}{2}\right]$}

Here, we evaluate the real representative $\Lambda^{(k)}$ of $\rho^{(k)}$ for different values of $k$ ($k=2,\,3,\ldots,\left[\frac{N}{2}\right]$) making use of 
Eqs. (\ref{elements}), (\ref{cg_explicit}),(\ref{Lk}).  
We then construct the real symmetric matrix $\Omega^{(k)}={\Lambda^{(k)}}\,G\left(\Lambda^{(k)}\right)^T$ for $k=2,\,3,\ldots,\left[\frac{N}{2}\right]$ and observe that 
$G\Omega^{(k)}=G{\Lambda^{(k)}}\,G\,{(\Lambda^{(k)})}^T$ has {\emph{non-degenerate eigenvalues}} $\lambda_0\neq\lambda_1\neq\lambda_2\neq\lambda_3$ when $k=2,3,\,\ldots,\left[\frac{N}{2}\right]$ and the highest eigenvalue $\lambda_0$ possesses an eigenvector $X_0$ satisfying the relation $X_0^T\,G\,X_0>0$.  The Lorentz canonical form $\widetilde{\Lambda}^{(k)}$, $k=2,\,3,\ldots,\left[\frac{N}{2}\right]$, is thus given by the diagonal matrix  (see (\ref{lambda1c})). 
\[
\widetilde{\Lambda}^{(k)}=\mbox{diag}\,\left(1,\,\sqrt{\lambda_1/\lambda_0},\,\sqrt{\lambda_2/\lambda_0},\,\pm\sqrt{\lambda_3/\lambda_0} \right).
\]
The eigenvalues $\lambda_\mu$, ($\mu=0,\,1,\,2,\,3$) of $G\Omega^{(k)}$ are dependent on the parameters `$a$', $k$ and $N$ characterizing the state $\vert D_{N-k,\,k}\rangle$, when 
$k$ takes any of the integral values greater than $1$ and less than $\left[\frac{N}{2}\right]$. Hence the canonical form $\widetilde{\Lambda}^{(k)}$, $k=2,3,\,\ldots,\left[\frac{N}{2}\right]$ is different for different states $\vert D_{N-k,\,k}\rangle$ unlike in the case of $\widetilde{\Lambda}^{(1)}$ (see (\ref{l32c})), the canonical form of W-class of states, which depends only on the number of qubits $N$. 

\section{Geometric representation of the states $\vert D_{N-k,k}\rangle$} 

In this section, based on the two different canonical forms of $\Lambda^{(k)}$ obtained in Sec.~III, we find the nature of canonical steering ellipsoids associated with the 
pure symmetric multiqubit states $\vert D_{N-k,k}\rangle$ belonging to SLOCC inequivalent families $\{{\mathcal{D}}_{N-k,\,k}\}$. To begin with, we give a brief outline~\cite{supra,can} of obtaining the steering ellipsoids of a two-qubit density matrix $\rho^{(k)}$ based on the form of its real representative $\Lambda^{(k)}$.

In the two-qubit state $\rho^{(k)}$, local projective valued measurements (PVM) $Q>0$, 	$Q=\sum_{\mu=0}^{3}\, q_\mu\, \sigma_\mu$,   $q_0=1$, $\sum_{i=1}^3\,q_i^2=1$ on Bob's qubit  leads to collapsed state of Alice's qubit characterized by its Bloch-vector 
${\mathbf p}_A=(p_1,\,p_2,\,p_3)^T$ through the transformation~\cite{supra} 
\begin{equation} 
\label{funda}
\left(1, p_1,\,p_2,\,p_3 \right)^T=\Lambda^{(k)}\,\left(1, q_1,\,q_2,\,q_3 \right)^T, \ \ q_1^2+q_2^2+q_3^2=1. 
 \end{equation}
Notice that the vector ${\mathbf{q}}_B=\left(q_1,\,q_2,\,q_3 \right)^T$, $q_1^2+q_2^2+q_3^2=1$ represents the entire Bloch sphere and the steered Bloch vectors ${\mathbf p}_A$ of Alice's qubit constitute an ellipsoidal surface ${\mathcal E}_{A\vert\,B}$ enclosed within the Bloch sphere. 
When Bob employs convex combinations of PVMs i.e., positive operator valued measures (POVMs), to steer  Alice's qubit, he can access the 
 points inside the steering ellipsoid. Similar will be the case when Bob's qubit is steered by Alice through local operations on her qubit. 

 For the Lorentz canonical form $\widetilde{\Lambda}^{(k)}_{I_c}$  (see (\ref{lambda1c})) of the two-qubit state $\widetilde{\rho}^{(k)}_{\,I_c}$, it follows from (\ref{funda}) that
\begin{equation}
		p_1=\sqrt{\frac{\lambda_1}{\lambda_0}}\,q_1, \ \ p_2=\sqrt{\frac{\lambda_2}{\lambda_0}}\,q_2,\ \ p_3=\pm \sqrt{\frac{\lambda_3}{\lambda_0}} q_3, \ \ 
					\end{equation} 
are steered Bloch points ${\mathbf {p}}_A$ of Alice's qubit.  They are seen to obey the equation 
		\begin{equation}
		\label{ellI}
		\frac{\lambda_0\, p_1^2}{\lambda_1}+ \frac{\lambda_0\, p_2^2}{\lambda_2}+ \frac{\lambda_0\, p_3^2}{\lambda_3}=1
		\end{equation}
		of an  ellipsoid with semiaxes  $(\sqrt{\lambda_1/\lambda_0}, \,\sqrt{\lambda_2/\lambda_0},\, \sqrt{\lambda_3/\lambda_0})$ and center $(0,0,0)$ inside the Bloch sphere $q_1^2+q_2^2+q_3^2=1$. We refer to this as the {\em canonical steering ellipsoid} representing the set of all two-qubit density matrices which are on the SLOCC orbit of the state $\widetilde{\rho}^{(k)}_{\,I_c}$ (see (\ref{rhokab})). 

For the second Lorentz canonical form $\widetilde{\Lambda}_{II_c}$ (see (\ref{lambda2c})), we get the coordinates of steered Alice's Bloch vector ${\mathbf{p}}_A$, on using (\ref{funda});
\begin{equation}
		\label{e2A}
	p_1=a_1q_1,\ \  p_2=-a_1q_2,\ \ p_3= \left(1-a_0\right)+a_0q_3, \   \ \ q_1^2+q_2^2+q_3^2=1
			\end{equation} 
and they satisfy the equation 
\begin{eqnarray}
		\label{sph}
	&& \frac{p_1^2}{a_1^2}+ \frac{p_2^2}{a_1^2}+ \frac{\left(p_3-(1-a_0)\right)^2}{a_0^2}=1.	
		\end{eqnarray}  
Eq. (\ref{sph}) represents the canonical steering spheroid (traced by Alice's Bloch vector ${\mathbf{p}}_A$) inside the Bloch sphere with its center at $(0,\,0,\, 1-a_0)$ and lengths of the semiaxes given by $a_0=\lambda_0/\phi_0$, $a_1=\sqrt{\lambda_1/\phi_0}$ given in (\ref{phi0}). In other words, a shifted spheroid inscribed within the Bloch sphere, represents  two-qubit states that are SLOCC equivalent to $\widetilde{\rho}^{(k)}_{II_c}$ (see (\ref{rho2})). 

\subsection{Canonical steering ellipsoids of W-class of states}

We have seen in Sec.~III B that the Lorentz canonical form of $\Lambda^{(1)}$, the real representative of the symmetric two-qubit state $\rho^{(1)}$ drawn from the W-class of states $\vert D_{N-1,1}\rangle$ has a {\emph{non-diagonal}} form (see (\ref{l32c}). On comparing (\ref{l32c}) with the canonical form in (\ref{lambda2c}), we get 
\begin{equation} 
a_1=\frac{1}{\sqrt{N-1}},\ \ a_0=\frac{1}{N-1}.
\end{equation}  
From (\ref{sph}) and the discussions prior to it,  it can be readily seen that the quantum steering ellipsoid associated with $\widetilde{\Lambda}^{(1)}$ in (\ref{l32c}) is a spheroid centered at $(0,0,\frac{N-2}{N-1})$ inside  the Bloch sphere, with fixed semiaxes lengths $(\frac{1}{\sqrt{N-1}},\, \frac{1}{\sqrt{N-1}},\, \frac{1}{N-1})$  (see Fig.~1). It is interesting to note that the Lorentz canonical form 
$\widetilde{\Lambda}^{(1)}$ is not dependent on the state parameter `$a$', $0\leq a<1$ and hence all states $\vert D_{N-1,\,1}\rangle$ in the family 
$\{{\mathcal{D}}_{N-1,\,1}\}$ are represented by a spheroid, all its parameters such as center, semiaxes, volume etc., dependent only on the number of qubits $N$.    
\begin{figure}[h]
	\begin{center}
		\includegraphics*[width=1.5in,keepaspectratio]{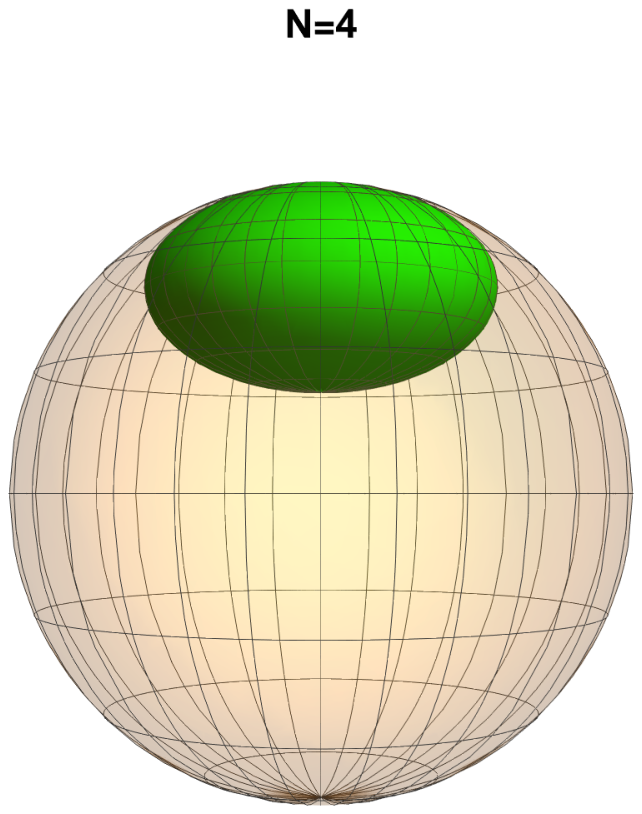}
		\includegraphics*[width=1.5in,keepaspectratio]{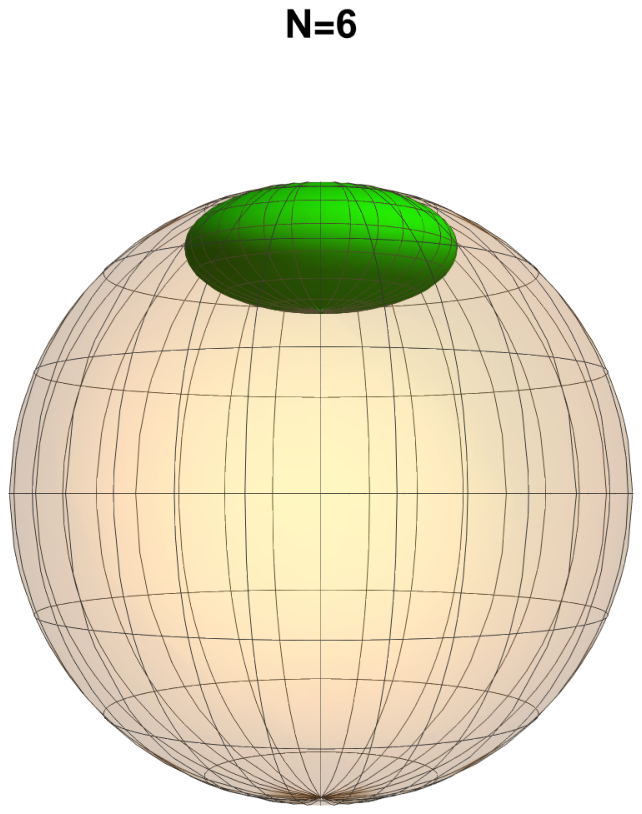}
		\includegraphics*[width=1.5in,keepaspectratio]{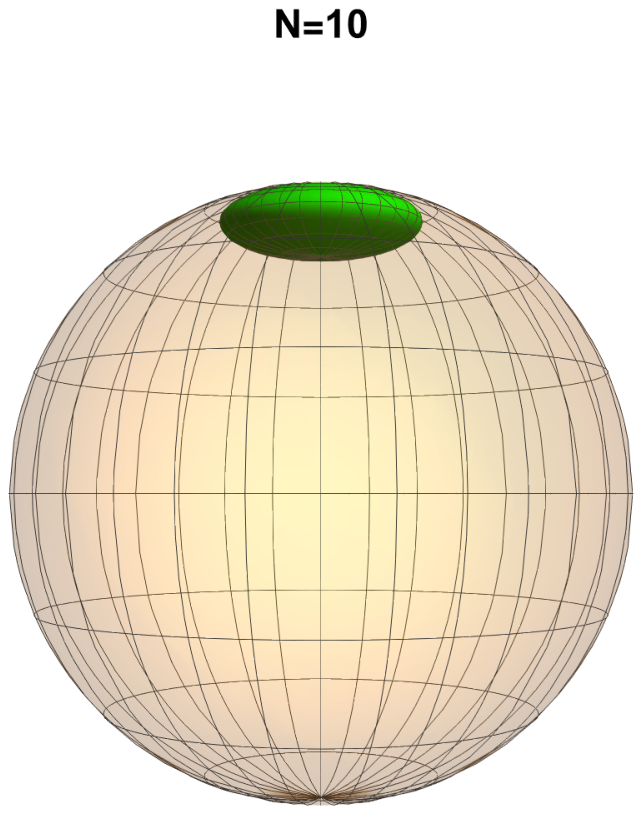}
		\includegraphics*[width=1.5in,keepaspectratio]{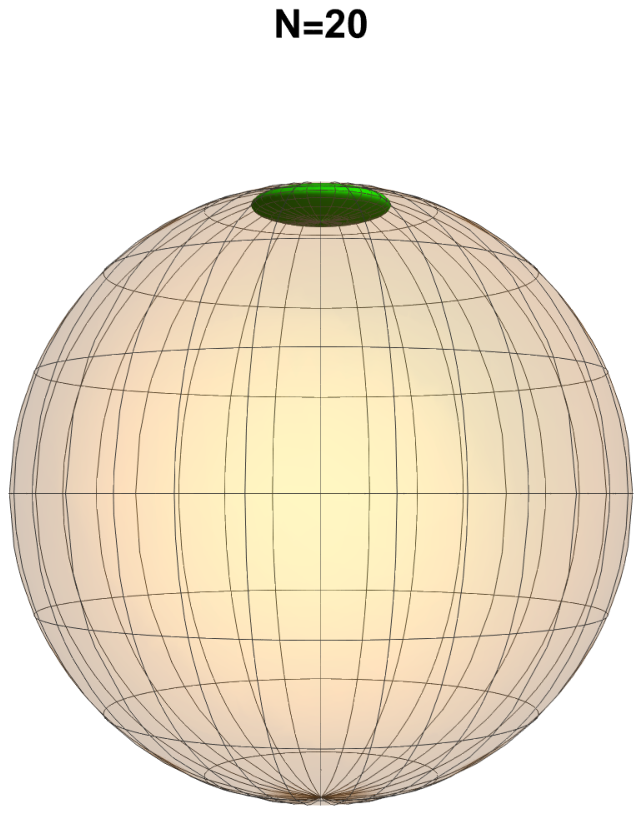}
		\caption{(Colour online) Steering spheroids inscribed within the Bloch sphere representing Lorentz canonical form $\widetilde{\Lambda}^{(1)}$ (see (\ref{l32c})) of W-class of states $\{{\mathcal{D}}_{N-1,\,1}\}$. The spheroids are centered  at $(0,0,\frac{N-2}{N-1})$ and the length of the semiaxes are given by $(\frac{1}{\sqrt{N-1}},\,\frac{1}{\sqrt{N-1}},\,\frac{1}{N-1})$.}
	\end{center}
\end{figure} 

\subsection{Canonical steering ellipsoids of the states  $\vert D_{N-k,k}\rangle$, $k=2,\,3,\ldots,\left[\frac{N}{2}\right]$}

As is seen in Sec.~III C, the Lorentz canonical form of $\Lambda^{(k)}$, $k=2,\,3,\ldots,\left[\frac{N}{2}\right]$, the real representative of the two-qubit states $\rho^{(k)}$ drawn from the pure symmetric $N$-qubit states $\vert D_{N-k,k}\rangle$, has the diagonal form (see (\ref{lambda1c})). The values of $\lambda_0$, $\lambda_1$, $\lambda_2$, $\lambda_3$, the eigenvalues of the matrix $G\,\Omega^{k}$ can be evaluated for each value of $k$, $k=2,\,3,\ldots,\left[\frac{N}{2}\right]$ for a chosen $N$. From (\ref{ellI}) and the discussions therein, it follows that the canonical steering ellipsoids of the states $\vert D_{N-k,k}\rangle$, $k=2,\,3,\ldots,\left[\frac{N}{2}\right]$ is an ellipsoid centered at the origin of the Bloch sphere with lengths of the semiaxes given by $\sqrt{\lambda_1/\lambda_0}$, $\sqrt{\lambda_2/\lambda_0}$, $\sqrt{\lambda_3/\lambda_0}$. The eigenvalues $\lambda_\mu$, $\mu=0,\,1,\,2,\,3$ of $G\Omega^{(k)}$ depend on the parameter `$a$' also, unlike in the case of W-class of states where they depend only on $N$, the number of qubits. Thus each state $\vert D_{N-k,k}\rangle$ belonging to the family $\{{\mathcal{D}}_{N-k,\,k}\}$, $k=2,\,3,\ldots,\left[\frac{N}{2}\right]$ is represented by an ellipsoid whose semiaxes depend on the values of $k$, $N$ and `$a$'.   The canonical steering ellipsoids corresponding to the $10$-qubit pure symmetric states $\vert D_{10-k,k}\rangle$ with chosen values of 
$k$ and `$a$'  are shown in In Fig.~2.
\begin{figure}[h]
	\begin{center}
		\includegraphics*[width=1.5in,keepaspectratio]{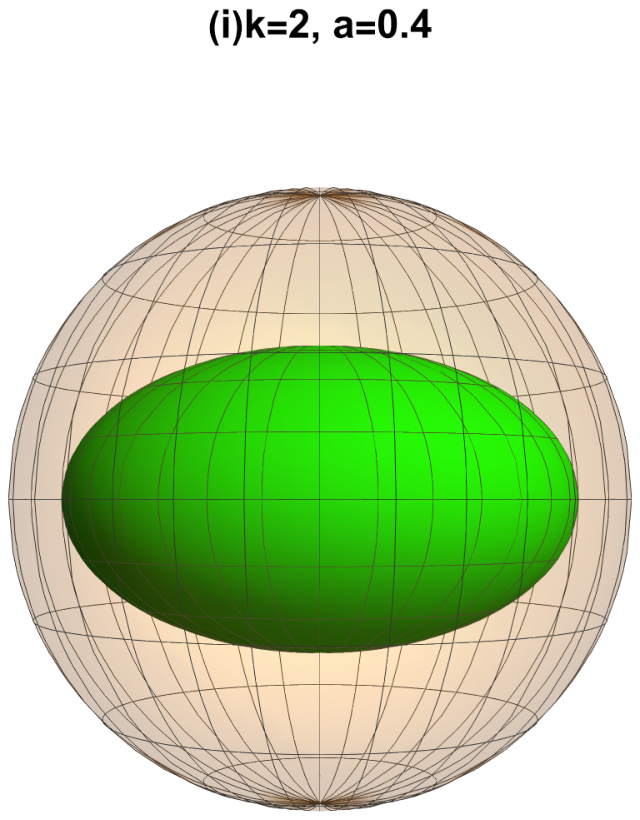}
		\includegraphics*[width=1.5in,keepaspectratio]{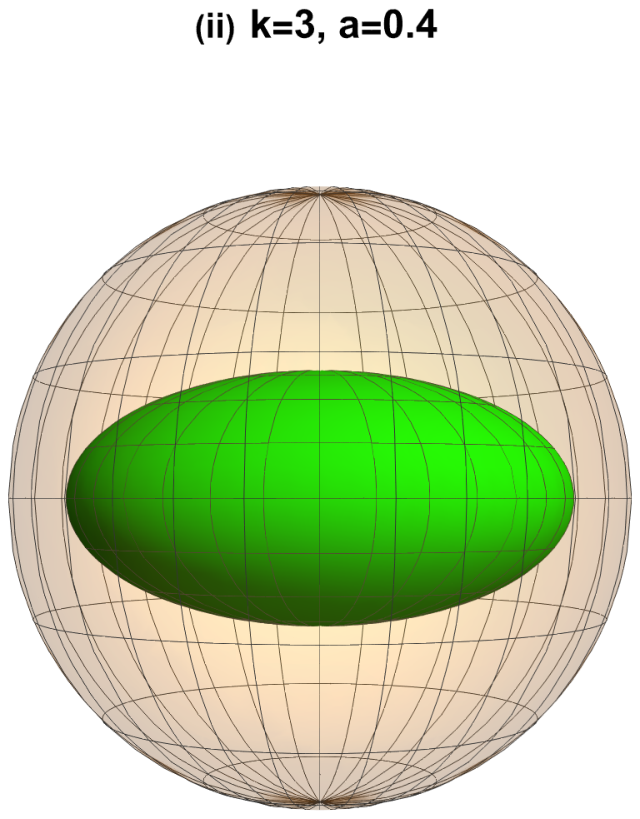}
		\includegraphics*[width=1.5in,keepaspectratio]{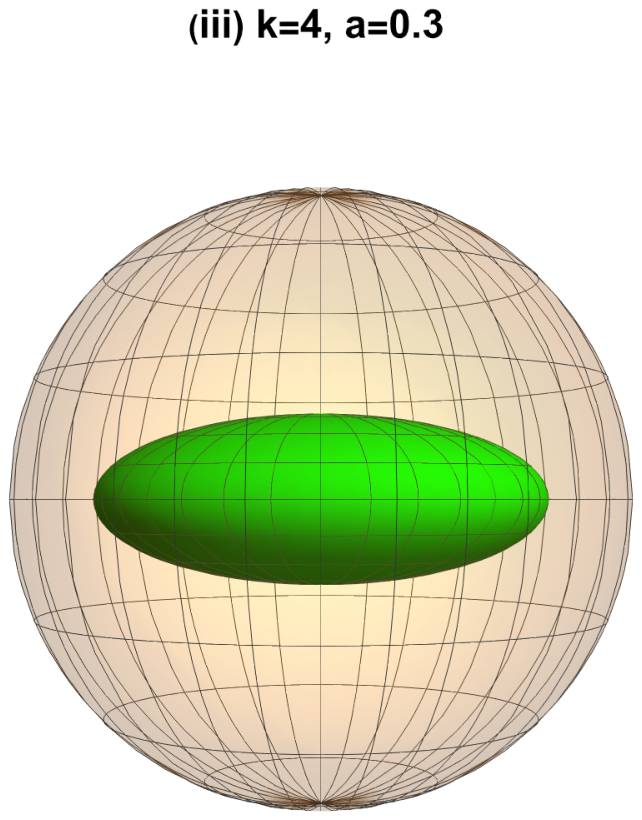}
		\includegraphics*[width=1.5in,keepaspectratio]{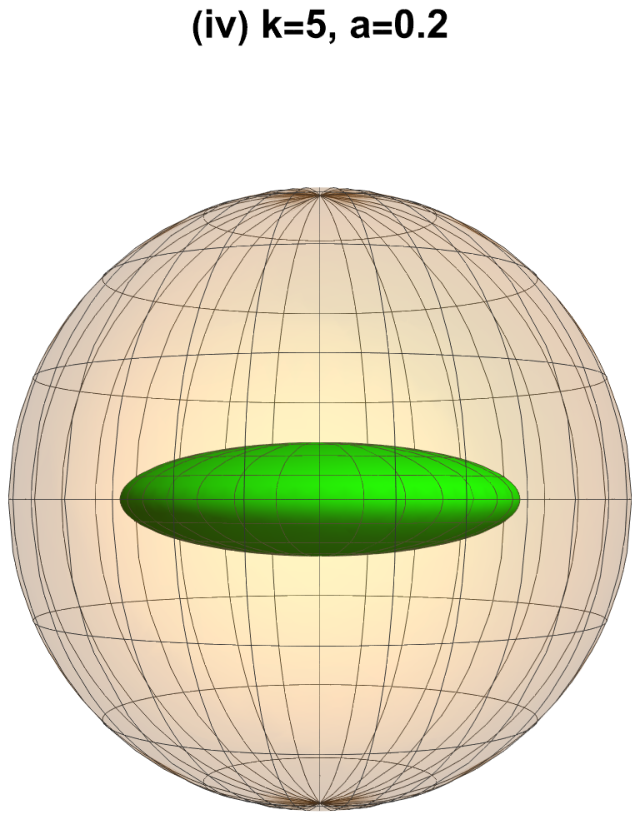}
		\caption{(Colour online) Steering ellipsoids centered at the origin of the Bloch sphere representing Lorentz canonical form of pure symmetric $10$-qubit states $\vert D_{10-k,k}\rangle$ for $k=1$ to $k=5$.
The length of the semi-axes of the ellipsoids for the $10$-qubit states chosen here are (i) $\left(0.91,\,0.71,\,0,62\right)$ 
  (ii) $\left(0.83,\,0.59,\,0.41\right)$
(iii) $\left(0.745,\,0.533,\,0.279\right)$  (iv)  $\left(0.656,\,0.53,\,0.185\right)$}.
	\end{center} 
\end{figure} 

In particular, the canonical steering ellipsoids corresponding to Dicke states are {\emph{oblate spheroids}} centered at the {\emph{origin}} (see Fig.3).
\begin{figure}[h]
	\begin{center}
		\includegraphics*[width=1.5in,keepaspectratio]{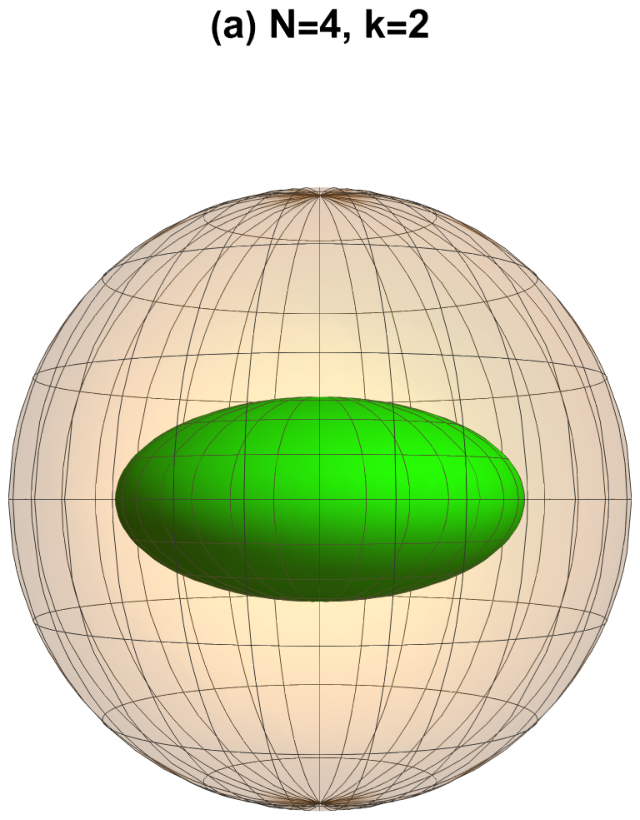}
		\includegraphics*[width=1.5in,keepaspectratio]{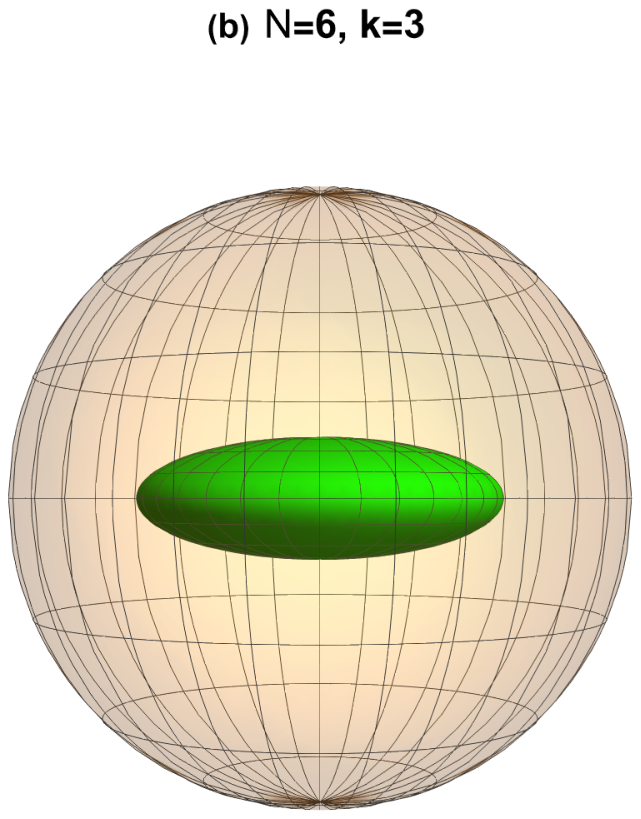}
		\includegraphics*[width=1.5in,keepaspectratio]{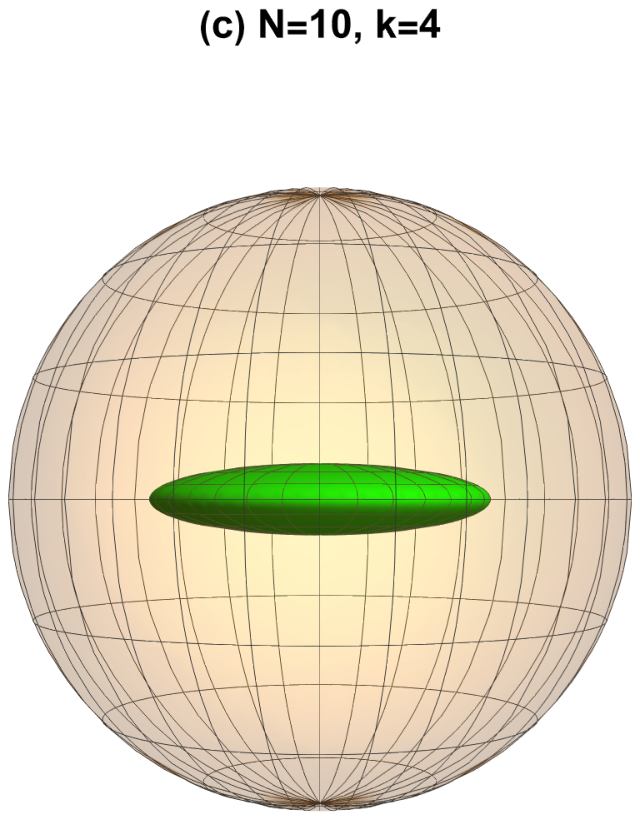}
		\includegraphics*[width=1.5in,keepaspectratio]{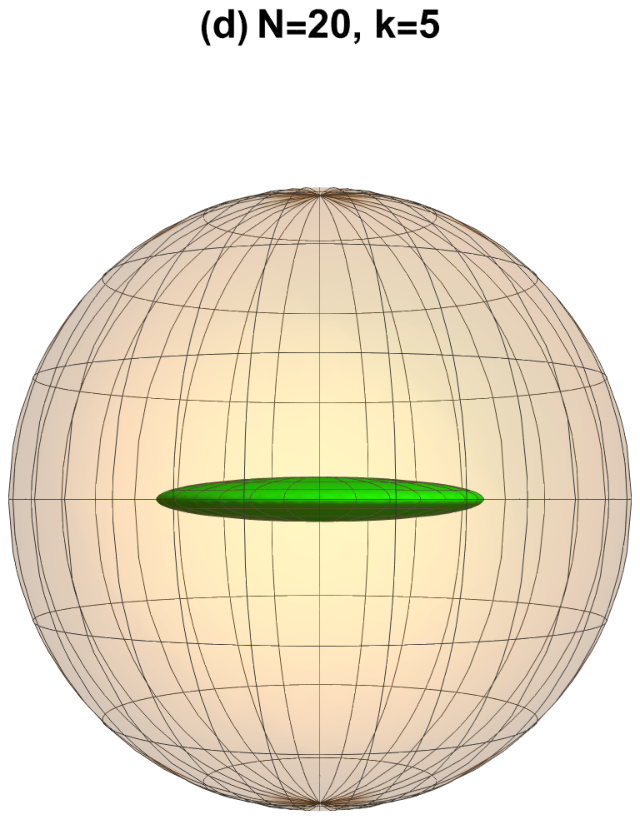}
		\caption{(Colour online) Oblate spheroids centered at the origin representing the Lorentz canonical form of the $N$-qubit Dicke states 
		$\left\vert N/2,N/2-k\right\rangle$ (equivalently, the states $\vert D_{N-k,k}\rangle$, with $a=0$)}
	\end{center}
\end{figure} 

\section{Volume monogamy relations for pure symmetric multiqubit states $\vert D_{N-k,k}\rangle$} 

 Monogamy relations restrict shareability of quantum correlations in a multipartite state. They find potential applications in ensuring security in quantum key distribution~\cite{Tehral,Paw}.   Milne {\em et. al.}~\cite{MilneNJP2014, MilnePRA2016} introduced a geometrically intuitive  monogamy relation for the volumes	of the steering ellipsoids representing  the two-qubit subsystems of multiqubit pure states, which is stronger than
	the well-known  Coffman-Kundu-Wootters monogamy relation~\cite{CKW}. 	
	In this section we explore  how volume monogamy relation~\cite{MilneNJP2014} imposes limits on volumes of the quantum steering ellipsoids representing the two-qubit subsystems $\rho^{(k)}=\mbox{Tr}_{N-2}\,\left[\vert D_{N-k, k}\rangle\langle D_{N-k, k}\vert\right]$ of pure symmetric multiqubit states $\vert D_{N-k, k}\rangle$.

For the two-qubit state $\rho_{AB}(=\rho^{(k)})$ (see (\ref{rho2q})), we denote by ${\mathcal E}_{A\vert\,B}$, the quantum steering ellipsoid containing all steered Bloch vectors of Alice when Bob carries out local operations on his qubit.  The volume of ${\mathcal E}_{A\vert\,B}$ is given by~\cite{jevtic2014}

\begin{equation} 
\label{xxx}
V_{A\vert B}=\left(\frac{4\pi}{3}\right)\, \frac{\vert\det \Lambda \vert}{(1-r^2)^2}, 
\end{equation}
where  $r^2=\mathbf{r}\cdot\mathbf{r}=r_1^2+r_2^2+r_3^2$ (see (\ref{ri})).  As the steering ellipsoid is constrained to lie within the Bloch sphere, i.e.,  $V_{A\vert B}\leq V_{\rm unit}=(4\pi/3)$,  one can choose to work with the {\emph {normalized volumes}} $v_{A\vert B}=\frac{V_{A\vert B}}{4\pi/3}$, the ratio of the volume of the steering ellipsoid to the volume of a unit sphere~\cite{MilnePRA2016}.  

The volume monogamy relation satisfied by a {\emph{pure}} three-qubit state  shared by Alice, Bob and Charlie is given by~\cite{jevtic2014,MilneNJP2014,MilnePRA2016} 
	\begin{equation}
	\label{vm}
	\sqrt{V_{A\vert B}} + \sqrt{V_{C\vert B}} \leq \sqrt{\frac{4\pi}{3}}. 
\end{equation} 
where $V_{A\vert B}, \ V_{C\vert B}$ are respectively the volumes of the ellipsoids corresponding to steered states of Alice and Charlie when Bob performs all possible local measurements on his qubit.  The {\emph {normalized}} form of the volume monogmay relation (\ref{vm}) turns out to be 
\begin{equation}
	\label{vmn}
	\sqrt{v_{A\vert B}} + \sqrt{v_{C\vert B}} \leq 1, 
\end{equation} 
 where  $v_{A\vert B}=\frac{V_{A\vert B}}{4\pi/3}$ are the {\emph{normalized volumes}}. 

The monogamy relation (\ref{vmn}) is not, in general, satisfied by mixed three-qubit states \cite{MilnePRA2016} and it has been shown 
that
\begin{equation}
	\label{vmnM}
	\left(v_{A\vert B}\right)^{\frac{2}{3}} + \left(v_{C\vert B}\right)^{\frac{2}{3}} \leq 1, 
\end{equation} 
is the volume monogamy relation for pure as well as mixed three-qubit states~\cite{MilnePRA2016}. 

As there are $\frac{1}{2}(N-2)(N-1)$ three-qubit subsystems in a $N$-qubit state, each of which obey monogamy relation (\ref{vmnM}), on adding these relations and simplifying,  one gets~\cite{MilnePRA2016} 
\begin{equation}
	\label{vmnN}
	\left(v_{A\vert B}\right)^{\frac{2}{3}} + \left(v_{C\vert B}\right)^{\frac{2}{3}}+\left(v_{D\vert B}\right)^{\frac{2}{3}}+\cdots \leq \frac{N-1}{2}. 
\end{equation} 
The relation (\ref{vmnN}) is the volume monogamy relation satisfied by pure as well as mixed $N$-qubit states~\cite{MilnePRA2016}. For $N=3$, it reduces to (\ref{vmnM}). 

For multiqubit states that are invariant under exchange of qubits, $v_{A\vert B}=v_{C\vert B}=v_{D\vert B}=\cdots=v_N$ where $v_N$ denotes the normalized volume of the steering ellipsoid corresponding to any of the $N-1$ qubits, the steering performed by, say $N$th qubit. Eq. (\ref{vmnN}) thus reduces to 
\begin{equation} 
\label{vmnNs}
(N-1) \left(v_N\right)^{\frac{2}{3}} \leq \frac{N-1}{2} \Longrightarrow \left(v_N\right)^{\frac{2}{3}} \leq \frac{1}{2} 
\end{equation}
implying that  $\left(v_N\right)^{\frac{2}{3}} \leq \frac{1}{2}$ is the volume monogamy relation for permutation symmetric multiqubit states.

\subsection{Volume monogamy relations governing the W-class of states $\{{\mathcal{D}}_{N-1,1}\}$} 
On denoting the normalized volume of a steering ellipsoid corresponding to the states $\vert D_{N-1,1}\rangle$ by $v^{(1)}_N$, we have (see (\ref{xxx}))  
\begin{equation} 
	\label{v32}
	v^{(1)}_N=\frac{\vert\det \Lambda^{(1)}\vert}{(1-r^2)^2}, 
\end{equation} 
where $\Lambda^{(1)}$ is given in (\ref{lambda32}) and 
\begin{equation}
	\label{r32}
	r_1=\frac{2a\sqrt{1-a^2}}{1+a^2(N-1)},\ \ r_2=0, \ \  r_3=1+\frac{2a^2}{1+a^2(N-1)}-\frac{2}{N}
\end{equation}
Under suitable Lorentz transformations, the real matrix $\Lambda^{(1)}$ (see (\ref{lambda32})) associated with the state $\rho^{(1)}$ (see (\ref{rho1_matrix}))  gets transformed to its Lorentz canonical form $\widetilde{\Lambda}^{(1)}$ 
(see (\ref{l32c})). It follows that (see (\ref{phi0}),  (\ref{ev32}))
 \begin{equation}
	\label{phi032}
\left(L_A\,\Lambda^{(1)}\, L^T_B\right)_{00}=\sqrt{\phi_0}=2\sqrt{N-1}\left[\frac{1-a^2}{N(1+(N-1)\,a^2)}\right]. 
\end{equation}
Using the property $\det L_A=\det L_B=1$ of orthochronous proper Lorentz transformations~\cite{KNS} and substituting $\vert\det\widetilde{\Lambda}^{(1)}\vert=\frac{1}{(N-1)^2}$ in (\ref{sl2c}), we obtain 
\begin{equation}
\label{int}
\vert\det\widetilde{\Lambda}^{(1)}\vert=\frac{1}{(N-1)^2}=\vert\det L_A\vert\, \vert\det L_B\vert   \left\vert\det\left(\frac{\Lambda^{(1)}}{\sqrt{\phi_0}}\right)\right\vert  
=\frac{ \vert\det\,\Lambda^{(1)}\vert}{\phi_0^2}.  
\end{equation}
Eq. (\ref{int}) leads to $\vert\det\,\Lambda^{(1)}\vert=\phi_0^2\,\vert\det\widetilde{\Lambda}^{(1)}\vert$. The normalized volume $v^{(1)}_N$ of the quantum steering ellipsoid corresponding to W-class of states thus becomes (see (\ref{v32}))
\begin{equation} 
\label{det32} 
v^{(1)}_N=\vert\det\widetilde{\Lambda}^{(1)}\vert\frac{\phi_0^2}{(1-r^2)^2}
\end{equation}
From (\ref{r32}) and (\ref{phi032}) it readily follows that $\phi_0^2=(1-r^2)^2$ and hence (see (\ref{det32})) the simple form for the normalized volume of the corresponding steering ellipsoid associated with  the two-qubit state $\rho^{(1)}$ turns out to be
\begin{eqnarray}
\label{vnW}
	v^{(1)}_{N}= \frac{\phi_0^2}{(N-1)^2\,(1-r^2)^2}=\frac{1}{(N-1)^2}. 
	\end{eqnarray} 
The volume monogamy relation $\left(v^{(1)}_{N}\right)^{\frac{2}{3}} \leq \frac{1}{2}$ (see (\ref{vmnNs})) takes the form 
  \begin{equation}
	\label{mnrW}
	\left(\frac{1}{(N-1)^2}\right)^{2/3}\leq \frac{1}{2} \, \Longrightarrow \,  (N-1)^{\frac{-4}{3}} \leq \frac{1}{2}
	\end{equation}
	and is readily satisfied for any $N\geq 3$ as can be seen in Fig.~4. 
	
\begin{figure}[h]
	\begin{center}
		\includegraphics*[width=3.5in,keepaspectratio]{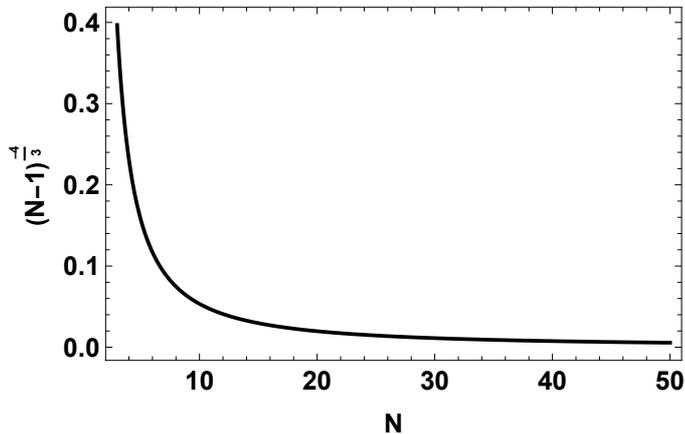}
	\caption{(Colour online) The LHS of the monogamy relation $(N-1)^{\frac{-4}{3}} \leq \frac{1}{2}$ is seen to be less than $\frac{1}{2}$ for the W-class of states $\vert D_{N-1,\,1}\rangle$ for any $N\geq 3$.}
	\end{center}
\end{figure} 	

\subsection{Relation between obesity of steering ellipsoids and concurrence} 

We recall here that the {\em obesity} ${\mathcal O}(\rho_{AB})=\vert \det\Lambda\vert^{1/4}$  of the quantum steering ellipsoid~\cite{MilneNJP2014} depicting a  
two-qubit state $\rho_{AB}$ is an upper bound for the concurrence $C(\rho_{AB})$:  
	 \begin{equation}
	 	\label{c&v}
	C(\rho_{AB})\leq {\mathcal O}(\rho_{AB}) =\vert \det\Lambda\vert^{1/4}.
	 \end{equation}
 Furthermore, if $\rho_{AB}\longrightarrow\widetilde{\rho}_{AB}=(A\otimes B)\rho_{AB}\, (A^\dag\otimes B^\dag)/({\rm Tr}(A^\dag\,A\otimes B^\dag B)\rho_{AB}]$, $A,B\in SL(2,C)$   it follows that~\cite{MilneNJP2014} 
 \begin{equation}
 	\label{cvratio}
 \frac{{\mathcal O}(\rho_{AB})}{C(\rho_{AB})}=\frac{{\mathcal O}(\widetilde{\rho}_{AB})}{C(\widetilde{\rho}_{AB})}.
 	\end{equation}
We make use of the relation (\ref{cvratio}) to obtain a relation for concurrence~\cite{Wootters} of a pair of qubits in the symmetric $N$-qubit pure states
$\vert D_{N-k,k}\rangle$, $k=1,\,2,\ldots,\left[\frac{N}{2}\right]$. For the states $\vert D_{N-1,1}\rangle$ belonging to W-class, we readily get  
(see (\ref{lambda32}), (\ref{l32c}))
 \begin{equation} 
\det\Lambda^{(1)}=\left(\frac{2(1-a^2)}{N(1+a^2 (N-1))}\right)^4, \ \ \ \det\widetilde{\Lambda}^{(1)}=\left(\frac{1}{N-1}\right)^2
\end{equation}
and thereby the obesities ${\mathcal O}(\rho^{(1)})$, 
${\mathcal O}(\widetilde{\rho}^{(1)})$: 
\begin{equation} 
\label{obwclass}
{\mathcal O}(\rho^{(1)})=\frac{2(1-a^2)}{N(1+a^2 (N-1))}, \ \ \ {\mathcal O}(\widetilde{\rho}^{(1)})=\frac{1}{\sqrt{N-1}}
\end{equation} 
It is not difficult to evaluate the concurrence of the canonical state $\widetilde{\rho}^{(1)}$ and it is seen that 
\begin{equation} 
\label{co}
C(\widetilde{\rho}^{(1)})={\mathcal O}(\widetilde{\rho}^{(1)})=\frac{1}{\sqrt{N-1}}.
\end{equation} 
We thus obtain (see (\ref{cvratio}),(\ref{co}))
\begin{eqnarray}
\label{cfin00}
C(\rho^{(1)})&=&{\mathcal O}(\rho^{(1)})=\frac{2(1-a^2)}{N(1+a^2 (N-1))}.
\end{eqnarray}	
The value of concurrence in (\ref{cfin00}) matches exactly with that obtained
using $C(\rho^{(1)})={\rm max} (0,\mu_1-\mu_2-\mu_3-\mu_4)$ where $\mu_1\geq\mu_2\geq\mu_3\geq \mu_4$ are square-roots of the eigenvalues of the matrix 
$R=\rho^{(1)}\,(\sigma_2\otimes\sigma_2)\, {\rho^{(1)}}^*\, (\sigma_2\otimes\sigma_2)$~\cite{Wootters}.
We have seen that the state	$\vert D_{N-1,\,1}\rangle$ reduces to W-state when $a=0$ and hence for the $N$-qubit W-state, 
concurrence of any pair of qubits is given by $C(\rho_W^{(1)})=\frac{2}{N}$ (see (\ref{cfin00})). 

\section{Summary} 
In this work, we have analyzed the canonical steering ellipsoids and volume monogamy relations of the pure symmetric $N$-qubit states characterized by two distinct Majorana spinors. We have shown that the entire W-class of states has a geometric representation in terms of a {\emph{shifted oblate spheroid}} inscribed within the Bloch sphere. 
The center of the spheroid, the length of its semiaxes and its volume are shown to be dependent only on the number of qubits $N$ implying that all states in the $N$-qubit W-class are characterized by a {\emph {single}} spheroid, shifted along the polar axis of the Bloch sphere.  All other SLOCC inequivalent families of pure symmetric $N$-qubit states with two distinct spinors are shown to be geometrically represented by {\emph {ellipsoids centered at the origin}}.  Except the W-state (and its obverse counterpart) which are represented by a {\emph{shifted spheroid}},  all other $N$-qubit Dicke states are represented by an {\emph{oblate spheroid centered at the origin}}.  A discussion on volume monogamy relations applicable to identical subsystems of a pure symmetric $N$-qubit state is given here and a volume monogamy relation applicable for W-class of states is obtained. A relation connecting concurrence of the two-qubit state and obesity of the associated quantum steering ellipsoid with its canonical counterparts is made use of to obtain concurrence of the states belonging to W-class. It would be interesting to examine the features of canonical steering ellipsoids and volume monogamy relations for the SLOCC inequivalent families of pure symmetric multiqubit  states with more than two distinct spinors; in particular, the class of pure symmetric $N$-qubit states belonging to GHZ-class (with three distinct spinors).

\section*{Acknowledgements}
BGD thanks IASC-INSA-NASI for the award of Summer Research Fellowship-2022, during this work. Sudha, ARU and IR are supported by the Department of Science and Technology (DST), India through Project No. DST/ICPS/QUST/2018/107.

	\end{document}